\documentclass[prb,twocolumn,amssymb]{revtex4}
\usepackage{graphicx}
\begin{document}
\title{Mott states under the influence of fermion-boson conversion: \\
Invasion of superfluidity }
\author{ Fei Zhou }
\affiliation{ Department of Physics and
Astronomy, University of British Columbia,\\ 6224 Agricultural
Road, Vancouver, British Columbia, Canada, V6T 1Z1}
\date{\today}
\begin{abstract}
I study the influence of fermion-boson conversion near Feshbach resonances 
on Mott 
states of Cooper pairs and demonstrate possible invasion of superfluidity.
The quantum dynamics of Fermi-Bose gases is studied using both an 
effective coupled $U(1)\otimes U(1)$ quantum rotor Hamiltonian and
a coupled XXZ $\otimes$ XXZ spin Hamiltonian. I also 
point
out two distinct branches of collective modes in superfluid states,
one of which involves anti-symmetric phase oscillations in fermionic and 
bosonic channels and is {\em always} gapped because of fermion-boson conversion.
\end{abstract}
\maketitle

Recent activities on atomic systems near Feshbach resonances have generated 
much 
interest. By varying the two-body scattering length near 
Feshbach resonances, several groups have 
successfully achieved fermionic superfluid in a strongly interacting 
regime\cite{Regal04,Zwierlein03,Jochim03,Hulet03}. Superfluid states 
near Feshbach
resonances turn out to be related to the BCS-BEC crossover studied long 
time 
ago\cite{Eagles69,Leggett80,Nozieres85,Melo93}. 
This was recently pointed out 
by a few 
groups\cite{Holland01,Timmermans01,Ohashi02,Stajic04,Falco04,Ho04}.
These groups have also made various efforts 
to incorporate the two-body resonance between Cooper pairs (open channel) 
and molecules (close channel)
explicitly in the many-body Hamiltonian
and clearly illustrated an interesting role of atomic resonances. 
Near a Feshbach resonance the usual Cooper 
pairing amplitude
and molecule condensate wave function are proportional to each other;
the Feshbach resonance introduces an effective interaction between 
fermions, the interaction constant of
which is $\gamma_{FB}^2/(2\mu-v)$. Here $\gamma_{FB}$ is the resonance width, 
$v$ is the detuning energy of
molecules and $\mu$ is the chemical potential of fermions.

In this letter I am going to study a Mott state under the influence of 
Feshbach resonances, especially the
effect of fermion-boson conversion.
A Mott state of bosons or of Cooper pairs appears whenever bosons or 
Cooper pairs in lattices are strongly repulsively 
interacting and if the corresponding filling factors are integers. 
One of important properties of a Mott state is its
incompressibility, or a finite energy gap in its excitation spectra. 
And when hopping is renormalized to be zero due to repulsive interactions, the 
number of 
particles at each site is strictly quantized and discrete; particles are 
{\em locally} conserved. 
I am going to show that in general Mott states are unstable 
with respect to fermion-boson conversion near Feshbach resonances.
The primary reason is that fermions or bosons involved in 
resonating conversion are not 
conserved separately. So the conversion  not only mediates an 
attractive interaction between fermions as realized before, but also, more importantly 
violates the local conservation law and
introduces {\em a novel mechanism to transport particles} in some limit.  
The necessary condition for particles to be localized,
or the number of fermions or bosons should be quantized might no longer exist. 
I find the invasion of superfluidity into a part of a parameter space 
where only Mott states are expected to be ground states if there are no Feshbach resonances.

The model I employ to study this subject is an M-orbit
Fermi-Bose 
Hubbard Model ({\em FBHM}). 
Consider the following general form of FBHM 

\begin{eqnarray}
&& H=H_f + H_b + H_{fb}; \nonumber \\
&& H_f=-t_f \sum_{<kl>, \eta,\sigma}
( f^\dagger_{k\eta \sigma} f_{l\eta\sigma} + h.c.)\nonumber \\
&& +\sum_{k, \eta, \sigma} (\epsilon_\eta -\mu) f^\dagger_{k\eta\sigma} 
f_{k\eta\sigma}
\nonumber \\
&& -\lambda \sum_{k,\eta,\xi} 
f^\dagger_{k\eta\uparrow}f^\dagger_{k\eta\downarrow}
f_{k\xi\downarrow}f_{k\xi\uparrow}
+\frac{V_f}{4} \sum_k \hat{n}_{fk} (\hat{n}_{fk}-1);
\nonumber \\
&& H_b=-t_b \sum_{<kl>} (b^\dagger_k b_l +h.c.)
\nonumber \\
&& +\sum_k (v-2\mu)b^\dagger_kb_k
+ V_b\sum_{k} \hat{n}_{bk} (\hat{n}_{bk}-1);
\nonumber \\
&& H_{bf}=-\gamma_{FB} \sum_{k,\eta}
(b^\dagger_{k} f_{k\eta\uparrow}f_{k\eta\downarrow}+h.c.)
\nonumber \\
&& +V_{bf} \sum_k \hat{n}_{bk} \hat{n}_{fk}.
\label{FBHM}
\end{eqnarray}
Here $k$, $\eta$ and $\sigma$ label lattice sites, on-site orbits and
spins; $\eta=1,2,...M$, $\sigma=\uparrow,\downarrow$.
$f^\dagger_{k\eta\sigma}$ $(f_{k\eta\sigma})$ is the creation(annihilation) 
operator
of a fermion at site $k$, with on-site orbital energy $\epsilon_\eta$ and spin 
$\sigma$.
$b^\dagger_k$($b_k$) is the creation 
(annihilation) operator of a boson at site $k$ (For simplicity, I assume there is 
only one bosonic orbital degree of freedom at each site).
The fermion and boson number operators are, respectively, 
$\hat{n}_{bk}=b^\dagger_kb_k$,$\hat{n}_{fk}
=\sum_{\eta,\sigma}f^\dagger_{k\eta\sigma}f_{k\eta\sigma}$. 
$t_f$ and $t_b$ are hopping integrals of fermions and bosons respectively and hopping
occurs over neighboring sites labeled as $<kl>$.
$\lambda$ is the attractive coupling constant in the Cooper channel which I 
assume to be much larger
than the rest of couplings. Finally, $V_f$, $V_b$ and $V_{fb}$ are
the strength of repulsive interactions between fermions, bosons, and 
fermions and bosons
in the density-density channel\cite{Petrov05}(One further assumes 
$V_{b}V_f > V^2_{fb}$
to ensure that homogeneous states are stable). 
FBHMs similar to Eq.(\ref{FBHM}) were previously applied
to study Bose-Fermi mixtures in optical lattices\cite{Albus03};
most recently an FBHM with fermion-boson conversion
was generalized to study the BCS-BEC crossover in lattices\cite{Carr05}. 
In the absence of the conversion term, FBHM consists of decoupled
(attractive) Fermi-Hubbard model and Bose-Hubbard model; main properties 
of latter are known \cite{Fisher89}.

I first study the {\em large-n} limit where $n_b$ and 
$n_f (\ll M)$, the average numbers of fermions and 
bosons are both much bigger than unity. Because $\lambda$ is much larger than other
coupling constants, the ground state of fermions for the on-site part 
of the 
Hamiltonian $H_f$ should be a BCS state. This suggests that it should be 
convenient to work with the following coherent state representation,

\begin{eqnarray}
&& |\{ \phi_{fk} \}; \{ \phi_{bk} \}>=
\prod_k 
\sum_{n_{bk}}
g_0(n_{bk})\frac{[\exp(-i\phi_{bk}) 
b_k^\dagger]^{n_{bk}}}{\sqrt{n_{bk}!}}
\nonumber \\
&& \otimes 
\prod_\eta (v_\eta + u_\eta 
\exp(-i\phi_{fk}) f^\dagger_{k\eta\uparrow} f^\dagger_{k\eta\downarrow}) 
|vac >.
\label{coherent}
\end{eqnarray}
Here $v_\eta, u_\eta$ are the coherence factor in the BCS wave function which
minizes the total on-site energy;
$g_0(n_{bk})$ is a unity for $n_{max}+n_b > n_{bk} > n_b - n_{max}$,
$n_{max}$ is much larger than one.
These states form a low energy Hilbert subspace
and are orthogonal in the limit which interests us,
or $<\{\phi'_{fk} \}; \{ \phi'_{bk} \}|$$
\{\phi_{fk} \}; \{ \phi_{fk}\}>$ is equal to zero if 
$\phi_{fk} \neq \phi'_{fk}$ or
$\phi_{bk} \neq \phi'_{bk}$.
At last, in the coherent-state representation one can show that
$\hat{n}_{fk}/2 =i \partial/\partial {\phi_{fk}}$,
and $\hat{n}_{bk}=i\partial/\partial {\phi_{bk}}$;
or

\begin{eqnarray}
&& [\frac{1}{2}\hat{n}_{fk}, \exp(-i\phi_{fk'})]= \delta_{k,k'} 
\exp(-i\phi_{fk}),
\nonumber \\
&& [\hat{n}_{bk}, \exp(-i\phi_{bk'})]= \delta_{k,k'} \exp(-i\phi_{bk}).
\label{conjugate}
\end{eqnarray}

In the subspace spanned by the coherent states defined above, I find that 
the 
effective Hamiltonian is

\begin{eqnarray}
&& H_{eff}=
-J_f \sum_{<kl>} \cos(\phi_{fk}-\phi_{fl})
+\frac{V_f}{4} \sum_{k} (\hat{n}_{fk} -n_f)^2 \nonumber \\
&&
-J_b \sum_{<kl>} \cos(\phi_{bk}-\phi_{bl})
+V_b \sum_{k} (\hat{n}_{bk} -n_b)^2 \nonumber \\
&&-\sum_{k} \Gamma_{FB} 
\cos(\phi_{fk}-\phi_{bk})
+V_{fb} (\hat{n}_{bk}-n_b) (\hat{n}_{fk}-n_f).
\label{effH}
\end{eqnarray}
The exchange couplings $J_f$, $J_b$ and $\Gamma_{FB}$ can be estimated
as
\begin{eqnarray}
&& J_f=t_f^2\sum_{\eta,\eta'}
\frac{u_\eta v_\eta u_\eta' v_\eta'}{E_\eta + E_\eta'}, \nonumber \\
&& J_b = n_b t_b, \Gamma_{FB}=\gamma_{FB} 
\sqrt{n_b}\sum_\eta u_\eta v_\eta;
\end{eqnarray}
$E_\eta=\sqrt{(\epsilon_\eta-\mu)^2 +\Delta_0^2}$ is the quasi-particle energy 
and $\Delta_0$ is the on-site BCS energy gap. 
Furthermore, $n_f$ and $n_b$ are functions of $\mu$, $v$ and $V_{f,b,fb}$:

\begin{eqnarray}
&& n_f=\frac{2V_b(\mu+V_f/4)-V_{bf}(2\mu-v+V_b)}{V_fV_b-V^2_{bf}}
\nonumber \\
&& n_b=\frac{V_f(2\mu-v+V_b)-V_{bf}(2\mu+V_f/2)}{2(V_fV_b-V^2_{bf})}.
\end{eqnarray}
Obviously, the detuning energy $v$ has to be sufficiently small in order for
the population of bosons $n_b$ to be positive. 

Eq.(\ref{conjugate}) 
and Eq.(\ref{effH}) 
define the low energy 
quantum dynamics of 
fermions and bosons under the influence of 
fermion-boson conversion near Feshbach 
resonances. 
In the absence of both Feshbach 
resonances ($\Gamma_{FB}=0$) and $V_{fb}$, the effective 
Hamiltonian describes two
decoupled sets of quantum $U(1)$ rotors in a lattice, the behaviors of 
which are well known. If $n_f/2$ or $n_b$ is a positive integer, the effective 
model can be used 
to study superfluid-Mott state transitions. 
A Mott phase corresponds to $U(1)$ symmetry restored states and
$U(1)$-symmetry breaking solutions represent a superfluid phase.
For the bosonic (Cooper pair) sector, the phase transition takes place when
$r_f=zJ_f/V_f$ ($r_b= zJ_b/V_b)$ is equal to a critical value 
$r_{fc}$($r_{bc}$) ($z(>1)$ is the coordination number).
The critical values which are of order of unity are usually 
calculated numerically.

In the presence of $\Gamma_{FB}$, the Hamiltonian describes a coupled
$U(1)\otimes U(1)$ quantum rotor model in a lattice. 
$U(1)\otimes U(1)$ symmetry breaking solutions
when both $r_{f,b}$ are much larger than unity correspond to a superfluid 
phase.
In the superfluid phase,
the Hamiltonian in Eq.(\ref{effH}) further leads to the following 
semiclassical equation of motion in the long wave length limit 

\begin{eqnarray}
&& \frac{\partial \phi_{fk}}{\partial t}=V_f \delta \hat{n}_{fk}
+ 2 V_{fb} \delta \hat{n}_{bk},\nonumber \\
&& \frac{\partial \phi_{bk}}{\partial t}= 2V_b \delta \hat{n}_{bk}
+ V_{bf} \delta \hat{n}_{fk},\nonumber \\
&& \frac{1}{2}\frac{\partial \delta{\hat{n}}_{fk}}{\partial t}= {J_f}
\Delta \phi_{fk}+\Gamma_{FB} (\phi_{bk}-\phi_{fk}),
\nonumber \\
&& \frac{\partial \delta{\hat{n}}_{bk}}{\partial t}={J_b}
\Delta \phi_{bk}+\Gamma_{FB} (\phi_{fk}-\phi_{bk}).
\end{eqnarray}
Here $\delta\hat{n}_{fk,bk}=\hat{n}_{fk,bk}-n_{f,b}$.
I have taken a continuum limit and $k$ labels the coordinate of
phases of bosons and fermion pairs ($\phi_{fk,bk}$) in this equation; 
$\Delta$ is a Laplacian operator.
The lattice constant has been set to be one.
This semiclassical equation shows that there are two branches of 
collective modes the dispersion relations of which are given below:

\begin{eqnarray}
&& a) \omega^2=\alpha |{\bf Q}|^2, 
\phi_f(\omega, {\bf Q} \rightarrow 0)=\phi_b(\omega, {\bf Q} \rightarrow 
0);
\nonumber \\ 
&& b) \omega^2=\Omega_0^2+\beta |{\bf Q}|^2, \nonumber \\ 
&& \phi_f(\omega, {\bf Q} \rightarrow 0)=-
\frac{V_f-V_{fb}}{V_b-V_{fb}}
\phi_b(\omega, {\bf Q}\rightarrow 
0).
\end{eqnarray}
$\phi_{f,b}(\omega,{\bf Q})$ are the Fourier components of phase
fields $\phi_{fk,bk}(t)$. 
In general, $\Omega_0$, $\alpha$ and $\beta$ depend on various parameters 
in the Hamiltonian; and $\Omega_0$ is always proportional to $\Gamma_{FB}$ and
$\alpha$ on the other hand is independent of $\Gamma_{FB}$.
When $V_{fb}=0$, $V_f=V_b=V_0$ and $J_f=J_b=J_0$,
the dispersion relations are given as $\Omega^2_0=2\Gamma_{FB} V_0$,
$\alpha=\beta=2J_0V_0$.

It is worth emphasizing that
in the long wave length limit,
mode a) is fully symmetric in phase oscillations of fermions and bosons, 
independent of various parameters;
mode b) represents out-of-phase oscillations in fermionic and bosonic 
channel and becomes fully antisymmetry when $V_b=V_f$.
In the absence of conversion ($\Gamma_{FB}=0$),
these two modes correspond to two gapless Goldstone modes 
associated with breaking two decoupled $U(1)$ symmetries.
However, in the presence of Feshbach resonances 
only the symmetric mode a) remains gapless 
corresponding to the usual Goldstone mode of superfluid
while the antisymmetric mode b)
is fully gapped because of the phase-locking effect of Feshbach 
resonances.


In general,
the wave functions for the 
many-body ground state and excitations $\Psi_n(\{ \phi_{bk} 
\}; \{ \phi_{fk} \})$($n=0$ is the ground state) are the eigenstates of the Hamiltonian
in Eq.(\ref{effH}). The boundary conditions are periodical along the directions
of $\phi_{fk,bk}$ with a period $2\pi$, so the wave functions are effectively 
defined on an 
$S^1\otimes S^1$ torus with radius of each $S^1$ equal to $2\pi$. 
If the average number $n_f/2$ and $n_b$ are integers, 
one introduces a {\em gauge} transformation
$\Psi \rightarrow \Psi \prod_k \exp(-i n_f\phi_{fk}/2-i n_b \phi_{bk})$;
the shifted number operators become
$\delta\hat{n}_{fk}/2=i\partial/\partial \phi_{fk}$,
$\delta\hat{n}_{bk}=i\partial/\partial \phi_{bk}$.
In this new basis, a spontaneous-symmetry breaking solution
with wave function $\Psi \sim \prod_k 
\delta(\phi_{fk}-\phi_0)\delta(\phi_{bk}-\phi_0)$
represents a typical superfluid state.
A symmetry-unbroken solution with wave function
$\Psi \sim \prod_k {(2\pi)^{-1}}$ $\exp(im_{fk}\phi_{fk})
\otimes \exp(im_{bk}\phi_{mk})$ ($m_{fk,bk}=0$ for all $k$) on the other hand 
corresponds to
a Mott state with $ \delta\hat{n}_{fk(bk)} \Psi=0$ or 
$\hat{n}_{fk(bk)}\Psi=n_{f(b)}\Psi$
at each lattice site.

To understand the influence of Feshbach 
resonances on Mott states, I now consider a situation where
again both $n_b$ and $n_f/2$ are {\em integers} and $r_f$ is much less than 
$r_{fc}$
so that Cooper pairs are in a Mott state in the absence of Feshbach resonance.
On the other hand $r_b$ is much less than the critical value $r_{bc}$ so 
that bosons are condensed.
For simplicity, I have also assumed that $V_{bf}$ is much smaller than $V_f$ so that
it can be treated as a perturbation.
I am interested in the responses of Cooper-pair Mott states to 
fermion-boson conversion  
and carry out the rest of discussions in a mean field approximation
({\em MFA}).
In this {\em MFA}, $\phi_{fk}=\phi_f$, $\phi_{bk}=\phi_b$ 
for any lattice site
k. The ground state
$\Psi_0(\phi_b, \phi_f)$ (again defined on an $S^1\otimes S^1$ torus with radius $2\pi$) is 
the lowest energy state of the 
following mean field Hamiltonian

\begin{eqnarray}
&& H_{MFA}=- V_f \frac{\partial^2}{\partial \phi^2_{f}}
-V_b \frac{\partial^2}{\partial \phi^2_{b}}
-2V_{fb} \frac{\partial}{\partial \phi_{f}}\frac{\partial}{\partial \phi_{b}}
\nonumber \\
&& -z (J_f  \Delta_f  \cos\phi_{f} + J_b  \Delta_b \cos\phi_{b} ) 
\nonumber \\
&& - \Gamma_{FB} \cos (\phi_{f}-\phi_{b}).
\label{MFA}
\end{eqnarray}
Here again $z$ is the coordination number; I have also introduced two self-consistent order parameters
\begin{equation}
\Delta_f= <\cos\phi_f >, 
\Delta_b=<\cos\phi_b >.
\label{OP0}
\end{equation}
Here $<>$ stands for an average taken in the ground state.
Notice that the order parameters defined above are nonzero only when the 
U(1) symmetries are 
broken; particularly, $\Delta_f$ is proportional to the usual BCS pairing 
amplitude. Following Eq.(\ref{coherent}) and discussions above one indeed 
shows that
$<f^\dagger_{k\eta\uparrow}f^\dagger_{k\eta\downarrow}> =(\sum_{\eta} u_\eta 
v_\eta) \Delta_f$,  $<b^\dagger_{k}>=\sqrt{n_b} 
\Delta_b$\cite{Zhou05}.
And $\Delta_{f,b}$ vanish in a Mott state and are nonzero for superfluid states.

As $z J_b$ is much larger than $V_b$, 
$\phi_b$ has very slow dynamics; and the corresponding ground state for $\phi_b$ 
can be approximated as a symmetry breaking solution.
That is

\begin{eqnarray}
&& \Psi_0(\phi_f, \phi_b)=\Psi_{0f}(\phi_f) \otimes \delta(\phi_b),
\nonumber \\
&& \Psi_{0f}(\phi_f)=\frac{1}{\sqrt{2\pi}}[1
+(\frac{z J_f}{V_f} \Delta_f 
+\frac{\Gamma_{FB}}{V_f}\Delta_b) \cos\phi_f].
\label{wf}
\end{eqnarray}
Here $\Delta_b$ should be approximately equal to one in this limit;
and in the zeroth order of $V_f^{-1}$, $\Psi_{0f}$ doesn't break the 
$U(1)$ symmetry and stands for a Mott-state solution.
Finally taking into account Eq.(\ref{OP0}) and (\ref{wf}),
one finds that the self-consistent solution 
to $\Delta_f$ is

\begin{equation}
\Delta_f= \frac{1}{2}\frac{\Gamma_{FB}}{V_f}
[1-\frac{1}{2}\frac{z J_f}{V_f}]^{-1}.
\label{order}
\end{equation}
In the absence of $\Gamma_{FB}$, $\Delta_f$ vanishes as expected for a Mott state.
However, the Mott state solution is unstable in the presence of any Feshbach 
conversion and the pairing order parameter $\Delta_f$ is always nonzero in this 
limit\cite{Zhou05a}.


I want to emphasize that the {\em average} number of fermion per site 
is not 
affected by the fermion-boson conversion and remains to be an even integer 
($2I$); rather, 
closely connected with the instability is the breakdown of particle quantization.
Indeed, one obtains in the {\em MFA} the following results for 
$\hat{n}_{fk}$,
$<\hat{n}_{fk}>=n_f=2I$, $<\delta^2 
\hat{n}_{fk}>=1/2 ({\Gamma_{FB}}/{V_f})^2 \approx 2 \Delta_f^2$.
This illustrates that the resonance between states with different numbers of 
fermions at a lattice site eventually leads to a nonzero pairing amplitude 
$\Delta_f$.

However, if $r_{b,f}$ are much smaller than $r_{bc,fc}$, then both Cooper pairs and
bosons are in Mott states. 
Similar calculations lead to
self-consistent solutions $\Delta_f=\Delta_b=0$ and more importantly, 
the following correlations
for $\Delta_{bf}^\pm$ $=<\cos(\phi_{b}\pm\phi_{f})>$\cite{Zhou05},

\begin{eqnarray}
\Delta_{bf}^-=\frac{\Gamma_{FB}}{2(V_f+ V_b)}, \Delta^+_{bf}=0.
\end{eqnarray}
The second equality above simply shows the absence of superfluidity.
But the first one indicates a subtle {\em hidden} order in the Mott states under 
consideration. Notice that $\Delta_{bf}^- $ $ \sim <b^\dagger_k 
f_{k\eta\uparrow}f_{k\eta\downarrow}>$ and represents a tri-linear order.

The main conclusions arrived here do not depend on the {\em large-n} 
approximation 
introduced here. One can consider the opposite limit by assuming $M=1$ and there 
is only one orbital degree of freedom at each lattice site.
The two interaction terms 
(with two interaction constant $\lambda$ and $V_f$) in $H_f$
(see Eq. (\ref{FBHM})) in the 
single 
orbital limit can be rewritten in one term: 
$V_f'/4 \sum_k \hat{n}_{fk}(\hat{n}_{fk}-1)$ if one identifies
$V_f'=V_f -2 \lambda$.
In the limit where $\lambda$ is much larger than $V_f$, fermions are paired  
at each lattice site. Furthermore I assume bosons have hard core 
interactions ($V_b=\infty$) such that there can be only zero or one 
boson at each site.

So the low energy Hilbert subspace ${\cal S}_k$ at each lattice site $k$ 
consists of four states: 1) no Cooper pair, no boson; 2) no Cooper pair, one boson;
3) one Cooper pair, no boson; 4) one Cooper pair, one boson.
They also correspond to a product of two pseudo spin $S=1/2$ 
subspaces: ${\cal S}_k={\cal S}_{fk}\otimes {\cal S}_{bk}$,
$ |\sigma^{z}_{fk}=\pm 1 > \in {\cal S}_{fk}$,
$ |\sigma^{z}_{bk}=\pm 1 > \in {\cal S}_{bk}$;
${\cal S}_k$ is the on-site Hilbert space, and
${\cal S}_{fk,bk}$ are the on-site pseudo 
spin spaces for fermions and bosons respectively.
$ |\sigma^z_{fk}=1>={f^\dagger_{k\uparrow}f^\dagger_{k\downarrow}}|vac>_f$,
$ |\sigma^z_{bk}=-1>=|vac>_f$; 
$ |\sigma^z_{bk}=1>=b^\dagger_k |vac>_b$,
$ |\sigma^z_{bk}=-1>=|vac>_b$.
$|vac>_{f,b}$ are the vaccuum of fermion and bosons respectively.
Finally, in this truncated subspace, the following identities hold
$\sigma^+_{fk}=f^\dagger_{k\uparrow}f^\dagger_{k\downarrow}$,
$\sigma^-_{fk}=f_{k\downarrow}f_{k\uparrow}$,
$\sigma^z_{fk}=2f^\dagger_{fk\uparrow}f^\dagger_{fk\downarrow}
f_{fk\downarrow}f_{fk\uparrow}-1$; 
$\sigma^+_{bk}=b^\dagger_{k}$,
$\sigma^-_{bk}=b_{k}$, $\sigma^z_{bk}=2b^\dagger_{k} b_k-1$.
So to have superfluidity, either $\sigma_{bk}$
or $\sigma_{fk}$, or both of them need to have a finite expectation value in the {\em XY} plane. 
For instance to have fermionic superfluid, the expectation values 
of $\sigma^\pm_{fk}$ need to be nonzero.

The effective Hamiltonian can then be written as
\cite{Zhou05}

\begin{eqnarray}
&& H^1_{eff}=
-J^1_b \sum_{<kl>} \sigma^x_{bk} \sigma^x_{bl} + \sigma^y_{bk}\sigma^y_{bl}
-h^z_b \sum_k \sigma^z_{fb} \nonumber \\
&& 
-J^1_f \sum_{<kl>} \sigma^x_{fk} \sigma^x_{fl} + \sigma^y_{fk}\sigma^y_{fl}
-\sigma^z_{fk}\sigma^z_{fl} -h^z_f \sum_k \sigma^z_{fk}  \nonumber \\
&& - \Gamma^1_{FB} \sum_{k} \sigma^x_{fk} \sigma^x_{bk} 
+ \sigma^y_{fk}\sigma^y_{bk}.
\end{eqnarray}
Here $J_f^1=t_f^2/V'_f$, $J^1_b=t_b$, $\Gamma^1_{FB}=\gamma_{FB}$, $h^z_f= \mu + 
V'_f/2$ and $h^z_b=\mu - v/2$.
The Hamiltonian is invariant under a rotation around the $z$-axis
or has an $XY$ symmetry.
The z-direction {\em fully} polarized phase of pseudo spins $\sigma_{bk}$ 
($\sigma_{fk}$) represents the Mott phase of bosons (fermions) and the $XY$ symmetry breaking 
states of pseudo spins $\sigma_{bk}$ ($\sigma_{fk}$) stand for the superfluid phase of bosons 
(fermions). The fermionic sector of this Hamiltonian was previously obtained and 
studied\cite{Emery76}; it was also used to study BEC-BCS crossover in lattices\cite{Carr05}.

When $\Gamma^1_{FB}=0$, the Mott phase of Cooper pairs with filling factor 
equal to 
one occurs when $h_{f}^z$ is much larger than $J_f$. Assume that in this case $h_b^z$ 
is much less than 
$J_b$ so that $\sigma_{bk}$ are ordered in the $xy$ plane; then bosons are condensed.
Taking into a finite amplitude of $\Gamma^1_{FB}$, one 
considers a solution where the pseudo spin symmetry of $\sigma_{bk}$
is spontaneously broken along a direction in the XY plane specified by $<{\bf \sigma}_{bk}>$
(the expectation value is taken in the ground state). In the molecular mean field 
approximation, the effective
external field acting on pseudo spins ${\bf \sigma}_{fk}$ is
${\bf h}_{f, eff}=z J_f <{\bf \sigma}_{fk}> +  \Gamma^1_{FB} <{\bf \sigma}_{bk}>$ $+ 
h^z_f {\bf e}_z$; 
and $<{\bf {\sigma}}_{fk}>$ is calculated self-consistently in the ground 
state when ${\bf h}_{f, eff}$ is applied.
One then arrives at the following solution 
$<{\bf \sigma}_{fk}> \cdot
<{\bf \sigma}_{bk}>$ $={\Gamma^1_{FB}}/{h^z_f}$
where $<{\bf \sigma}_{fk}>$ has been projected along the 
direction of $<{\bf \sigma}_{bk}>$ which lies in the $XY$ plane. As mentioned before,
development of such a component signifies superfluidity, or nonzero pairing 
amplitude.

In conclusion, I have shown that certain Mott states are unstable with respect 
to the resonating fermion-boson conversion;
in general superfluidity invades Mott phases because of the fermion-boson 
conversion. 
This work is supported by a Discovery grant from NSERC, Canada and a research grant from UBC.
I acknowledge useful discussions with I. Affleck, E. Cornell and R. Hulet.


\begin{thebibliography}{99}

\bibitem{Regal04}
M. Greiner, C. A. Regal, and D. S. Jin, Nature {\bf 426}, 537 (2003);
C. A. Regal, M. Greiner, and D. S. Jin, Phys. Rev. Lett. {\bf 92}, 040403 (2004).
\bibitem{Jochim03}
S. Jochim et al., Science {\bf 302}, 2101 (2003).
\bibitem{Zwierlein03} M. W. Zwierlein et al., Phys. Rev. Lett. {\bf 91}, 250401 (2003).
\bibitem{Hulet03}K. E. Strecker, G. B. Partridge and R. G. Hulet,
Phys. Rev. Lett. {\bf 91}, 080406 (2003).
\bibitem{Eagles69} D. M. Eagles, Phys. Rev. {\bf 186}, 456 (1969).
\bibitem{Leggett80} A. J. Leggett, J. Phys. C (Paris) {\bf 41}, 7 (1980)
\bibitem{Nozieres85} P. Nozieres and S. Schmitt-Rink, J. Low. Temp. Phys. {\bf 59}, 195 (1985).
\bibitem{Melo93}C. S. De Melo, M. Randeria, and J. Engelbrecht, Phys. Rev. Lett.
{\bf 71}, 3202 (1993).
\bibitem{Holland01} M. Holland {\em et al.}, Phys. Rev. Lett. {\bf 87}, 120406 
(2001).
\bibitem{Timmermans01} E. Timmermans {\em et al.}, Phys. Lett. A {\bf 285}, 228 
(2001).
\bibitem{Ohashi02}Y. Ohashi and A. Griffin, Phys. Rev. Lett. {\bf 89}, 130402 
(2002).
\bibitem{Stajic04}J. Stajic {\em et al.}, Phys. Rev. A {\bf 69}, 063610 (2004).
\bibitem{Falco04} G. M. Falco and H. T. C. Stoof, Phys. Rev. Lett. {\bf 92}, 140402 (2004).
\bibitem{Ho04}
R. B. Diener and T.L. Ho, cond-mat/0404517.
\bibitem{Petrov05} The atom-molecule and molecule-molecule scattering 
lengths have been calculated previously.
For instance, D. S. Petrov, C. Salomon, and G. V. Shlyapnikov,
Phys. Rev. A {\bf 71}, 012708 (2005) and references therein.
\bibitem{Albus03} In optical lattices, Bose-Fermi mixtures 
without fermion-boson conversion were studied in A. Albus {\em et al.}, Phys. 
Rev. A {\bf 68}, 023606 (2003); M. Lewenstein {\em et al.}, Phys. Rev. Lett. 
{\bf 92}, 050401 (2004); M. Cramer {\em et al.}, Phys. Rev. Lett. {\bf 93}, 
190405 (2004); D. W. Wang {\em et al.}, cond-mat/0410494.
\bibitem{Carr05} L. D. Carr and M. J Holland, cond-mat/0501156.
\bibitem{Fisher89} M. P. A. Fisher, P. B. Weichman, G. Grinstein, and D. S. Fisher, Phys. Rev.
{\bf B 40}, 546 (1989); also discusions on Bose-Hubbard physics in optical 
lattices, D. Jaksch {\em et al.}, Phys. Rev. Lett. {\bf 81}, 3108 (1998).
Bosonic Mott states were observed recently, see M. Greiner {\em et al.}, Nature
(London) {\bf 415}, 39 (2002).
\bibitem{Zhou05}F. Zhou, unpublished (2005). 
\bibitem{Zhou05a} Eq.(\ref{order}) implies the 
delocalization of Cooper pairs due to fermion-boson conversion, which can be 
shown explicitly.
This result also indicates that the pairing amplitude be strongly renormalized by repulsive interactions
in the density-density channel, an alternative point of view which is substantiated in Ref.\onlinecite{Zhou05}.  
\bibitem{Emery76} V. J. Emery, Phys. Rev. B {\bf 14}, 2989 (1976).










\end{thebibliography}
\end{document}